\documentstyle[11pt]{article}

 \csname
@addtoreset\endcsname{equation}{section}

\def\a{\alpha} \def\ad{\dot{\a}} \def\ua{{\underline \a}}
\def\b{\beta}  \def\bd{\dot{\b}} \def\ub{{\underline \b}}
\def\c{\gamma} 

\def\d{\delta} 

\def\e{\epsilon} \def\he{\hat{\epsilon}}

\def\f{\phi}
\def\F{\Phi}

\def\k{\kappa}
\def\l{\lambda}
\def\L{\Lambda}
\def\m{\mu}
\def\n{\nu}
\def\r{\rho}
\def\s{\sigma}

\def\t{\tau}

\def\o{\omega}
\def\pb{{\bar\pi}}

\def\cF{{\cal F}}

\def\cR{{\cal R}}

\def\cA{{\cal A}}
\def\hcA{\hat\cA}
\def\hF{\hat{\F}}
\def\hcF{\hat{\cal F}}

\def\bp{{\bar \pi}}

\def\yb{{\bar y}}
\def\zb{{\bar z}}
\def\vb{{\bar v}}
\def\uba{{\bar u}}

\def\ra{\rightarrow}

\def\del{\partial}
\def\na{\nabla}
\let\la=\label
\let\bm=\bibitem
\def\nn{\nonumber}
\newcommand{\eq}[1]{(\ref{#1})}

\newcommand{\w}[1]{\\[0.#1cm]}
\def\eqs#1#2{(\ref{#1}-\ref{#2})} 

\def\be{\begin{equation}}
\def\ee{\end{equation}}
\def\bea{\begin{eqnarray}}
\def\eea{\end{eqnarray}}
\def\ba{\begin{array}}
\def\ea{\end{array}}

\def\ft#1#2{{\textstyle{{\scriptstyle #1}
\over {\scriptstyle #2}}}}

\def\ed{\end{document}}

\newcommand{\hoch}[1]{$\, ^{#1}$}

\newcommand{\tamphys}{\it\small Center for Theoretical Physics, Texas
A\&M University, College Station, TX 77843, USA}

\newcommand{\groningen}{\it\small Institute for Theoretical Physics,
Nijenborgh 4, 9747 AG Groningen,The Netherlands}

\newcommand{\auth}{\large E. Sezgin \hoch{1\dagger} and  P. Sundell
\hoch{2\star} }

\begin{document}

\hfill{CTP-TAMU-39/99}

\hfill{UG/23-00}

\hfill{hep-th/0012168}

% \hfill{\today}
\vspace{20pt}
\begin{center}

%%%%%%%%%%%%%%%%%%%%%%%%%%%%%%%%%%%%%%%%%%%%%%%%%%%%%%%%%%%%%%%%%%%%%%%%%

{\Large \bf On Curvature Expansion of Higher Spin Gauge Theory}

%%%%%%%%%%%%%%%%%%%%%%%%%%%%%%%%%%%%%%%%%%%%%%%%%%%%%%%%%%%%%%%%%%%%%%%%%

\vspace{30pt}

%\authookmarks

\auth

\vspace{15pt}

\begin{itemize}

\item[$^1$] \tamphys
\item[$^2$] \groningen

\end{itemize}

\vspace{30pt}

{\bf Abstract}

\end{center}

We examine the curvature expansion of a the field equations of a
four--dimensional higher spin gauge theory extension of anti-de
Sitter gravity. The theory contains massless particles of spin
$0,2,4,...$ that arise in the symmetric product of two spin $0$
singletons. We cast the curvature expansion into manifestly
covariant form and elucidate the structure of the equations and
observe a significant simplification.

% {\vfill\leftline{}\vfill \vskip 10pt \footnoterule {\footnotesize
% \hoch{\dagger} Research supported in part by NSF Grant PHY-0070964
% and by Stichting FOM, Netherlands. \vskip -12pt}

{\vfill\leftline{}\vfill \vskip 10pt \footnoterule {\footnotesize
To be published in the
proceedings of the G\"{u}rsey Memorial Conference II ``{\sl M-theory
and dualities}'', Istanbul, June 2000. \\
\hoch{\dagger} Research supported in part by NSF Grant PHY-0070964. \\
\hoch{*} Research supported in by Stichting FOM. \vskip -12pt}

\pagebreak

\setcounter{page}{1}

%%%%%%%%%%%%%%%%%%%%%%%%%%%%%%%%%%%%%%%%%%%%%%%%%%%%%%%%%%%%%%%%%%%%%

\section{Introduction}

%%%%%%%%%%%%%%%%%%%%%%%%%%%%%%%%%%%%%%%%%%%%%%%%%%%%%%%%%%%%%%%%%%%%%

It is reasonable to assume that the interactions of quantum
gravity simplifies in the limit of high energies, such that only a
limited spectrum survives whose interactions are governed by a
symmetry group, analogously to how supergravity/supersymmetry
emerges in the low energy limit. Clearly, we do not expect this
theory to be another supergravity theory, but instead it seems
much more suggestive to consider some massless higher spin
extension of supergravity. These are gauge theories based on
infinite dimensional algebras which are essentially given by the
enveloping algebras of an underlying anti-de Sitter superalgebra.
Thus the resulting higher spin gauge theory is an extension of the
corresponding gauged supergravity theory, capable of interpolating
between supergravity at low energies and its higher spin extension
at high energies.

The higher spin gauge theories in four dimensions have been primarily
developed by Vasiliev \cite{v}. For reviews, see \cite{vr,ss}. In
this report we shall describe some of the basic properties of the
higher spin theories in the context of an ordinary
nonsupersymmetric algebra \cite{kv}. We also examine the curvature
expansion of the field equations and write these on a manifestly
covariant form (i.e. without reference to the special anti-de
Sitter solution). In particular, we point at a cancellation of certain structures at
higher orders (beginning at the second order) that leads to a significant
simplification of the higher spin equations. The covariant form of the
action for the physical gauge fields, but
with the lower spin sector set equal to zero, was
given up to cubics in curvatures already in \cite{vc}, while the covariant
field equations given here include all physical fields as well as
auxiliary gauge fields and are valid to arbitrary order in curvatures.

The paper is organized as follows. In Section 2 we review the
formulation of anti-de Sitter gravity in the constraint formalism,
the bosonic higher spin algebra and its unitary representation on
a spin zero singleton, and give the corresponding basic field
content of the higher spin gauge theory. With these preliminaries,
we then discuss Vasiliev's procedure for constructing interactions
in Section 3. In Section 4 we examine the resulting covariant
curvature expansion of the higher spin equations. In Section 5 we
conclude and describe briefly work in progress based on the
results on Section 4.

%%%%%%%%%%%%%%%%%%%%%%%%%%%%%%%%%%%%%%%%%%%%%%%%%%%%%%%%%%%%%%%%%%%%%

\section{Preliminaries}

%%%%%%%%%%%%%%%%%%%%%%%%%%%%%%%%%%%%%%%%%%%%%%%%%%%%%%%%%%%%%%%%%%%%%

Our starting point is the $D=4$ Einstein's equation with a
negative cosmological constant:

\be
R_{\m\n}-\ft12 R g_{\m\n}+\L g_{\m\n}=0\quad ,\qquad \L<0
\la{ee}
\ee

Viewed as a curvature constraint it leaves ten components of the
Riemann tensor $R_{\m\n,\l\r}$ unconstrained. These form an
irreducible tensor called the Weyl curvature tensor. In van der
Waerden notation
%%%%%%%%%%%%%%%%%%%%%%%%%%%%%%%%%%%%%%%%%%%%%%%%%%%%%%%%%%%%%%%%%%%%
\footnote { An $SO(3,1)$ vector $V_a=(\s_a)^{\a\ad}V_{\a\ad}$
where $\a$ and its hermitian conjugate $\ad$ are two component
indices raised and lowered by $\e_{\a\b}=\e^{\a\b}$ using NE-SW
see--saw rules for both dotted and undotted indices.}
%%%%%%%%%%%%%%%%%%%%%%%%%%%%%%%%%%%%%%%%%%%%%%%%%%%%%%%%%%%%%%%%%%%%
this tensor is $C_{\a\b\c\d}$. We can write \eq{ee} in an
equivalent first order form as follows:

\bea
{\cR}_{\m\n,ab}&=& e_\m{}^c e_\n{}^d
(\s_{cd})^{\a\b}(\s_{ab})^{\c\d} C_{\a\b\c\d}+ {\rm h.c.}\ ,
\la{weyl}\w2 \cR_{\m\n,a}&=&0\ ,
\la{tor}
\eea

where $\cR=d\o+g\o\wedge \o={1\over 2i}(\cR^{ab}M_{ab}+ 2 \cR^a P_a)$ is
$SO(3,2)$ valued and $g$ is gauge coupling. The $SO(3,2)$ gauge
field has components $\o_\m{}^{ab}$ and $\o_\m{}^a=\sqrt{2}\kappa^{-1}
e_\m{}^a$, that we identify with the Lorentz connection and the
vierbein, respectively. Here $\kappa^2$ is the 4D Newton's
constant. The Lorentz valued curvature is related to the ordinary
Riemann tensor through

\be
\cR_{\m\n,ab}= R_{\m\n,ab} +4 g \kappa^{-2}e_\m{}^{[a}
e_\n{}^{b]}\ ,
\ee

while the $\cR_{\m\n,a}$ is the usual torsion. Tracing \eq{weyl}
with $e^{\n,b}$ gives ${\cR}_{\m a}=0$, which together with
\eq{tor} yields \eq{ee} with the identification

\be
\L=-{6g^2\over \kappa^2}\ .
\la{cc}
\ee

In order to extend $SO(3,2)$ to a higher spin algebra we start
from the following realization of $SO(3,2)$ in terms of Grassmann even
$SO(3,2)$ spinors:

\bea
M_{ab}&=& \ft14 (\s_{ab})^{\a\b}y_\a y_\b+\ft14 (\bar{\s}_{ab})^{\ad\bd}
\yb_{\ad}\yb_{\bd}\ ,  \la{osc1}
\w2
P_a&=& \ft12(\s_a)^{\a\ad}y_\a\yb_{\ad}\ ,
\la{osc2}
\eea

where the spinors satisfy the following oscillator algebra
%%%%%%%%%%%%%%%%%%%%%%%%%%%%%%%%%%%%%%%%%%%%%%%%%%%%%%%%%%%%%%%%%%
\footnote {The spinor $Y_{\ua}\equiv (y_\a,\yb_{\ad})$ is a
Majorana spinor of $S0(3,2)\simeq Sp(4)$. The algebras
\eqs{osc3}{osc4} and \eqs{eosc1}{eosc4} can be written in a
manifestly $Sp(4)$ invariant form; e.g. $Y_{\ua}\star
Y_{\ub}=Y_{\ua}Y_{\ub}+C_{\ua\ub}$.}
%%%%%%%%%%%%%%%%%%%%%%%%%%%%%%%%%%%%%%%%%%%%%%%%%%%%%%%%%%%%%%%%%:

\bea
y_\a\star y_\b&=& y_\a y_\b+i\e_{\a\b}\ ,\quad \yb_{\ad}\star \yb_{\bd}=
\yb_{\ad} \yb_{\bd}+i\e_{\ad\bd}\ ,
\la{osc3}
\w2
y_\a\star \yb_{\bd}&=& y_\a \yb_{\bd}\ ,\qquad\qquad \yb_{\ad}\star y_\b=
\yb_{\ad} y_\b\ .
\la{osc4}
\eea

Here $\star$ denotes the operator product and the products
without $\star$ are Weyl ordered (i.e. totally symmetric)
products. The $\star$ product of two arbitrary functions of $y$ and $\yb$ is
given by

\be
F\star G=\int d^4u~ d^4v ~ F(y+u,\yb+\uba) ~G(y+v,\yb+\vb) ~e^{i(u_\a
v^\a+\uba_{\ad}\vb^{\ad})}\ .\la{sp}
\ee

An irreducible highest weight representation $D(E,s)$ of $SO(3,2)$ is
labeled by the energy $E$ and spin $s$ of its ground state, where the energy
and spin operators are $M_{05}$ and $M_{12}$, respectively. Unitarity requires
$E\geq s+1$, $(E,s)=(\ft12,0)$ or $(E,s)=(1,\ft12)$. In the case that $s\geq
1$, there are extra null-states for $E=s+1$, and $D(s+1,s)$,
$s\geq 1$, are thus referred to as massless representations. The
scalar $D(1,0)$, the pseudo--scalar $D(2,0)$ and the
spin--$\ft12$ representation $D(\ft32,\ft12)$ are also referred
to as massless representations. The two special cases
$D(\ft12,0)$ and $D(1,\ft12)$ have only a finite number of states
of any given spin and cannot propagate in four dimensions. These
are the Rac and Di singletons, whose dynamics is restricted to
the conformal boundary of AdS. The states in the Fock space of
the oscillators \eqs{osc3}{osc4} with even and odd occupation
number generate the weight space of $D(\ft12,0)$ and
$D(1,\ft12)$, respectively.

In this report we shall consider a four--dimensional bosonic
higher spin extension of $SO(3,2)$ obtained from the algebra of
polynomials $P(y,\yb)$ in $y_\a$ and $\yb_{\ad}$ modulo the
following projection and reality conditions:

\be
\t(P(y,\yb))\equiv P(iy,i\yb)=-P\quad ,\qquad  P^{\dagger}=-P\ .
\ee

These conditions define the Lie algebra $hs_2(1) $ \cite{kv} with respect to the
bracket $[P,Q]=P\star Q-Q\star P$, since $\t$ and the $\dagger$
have the properties:

\be
\t(P\star Q)=\t(Q)\star \t(P)\ ,\quad (P\star Q)^{\dagger}= Q^{\dagger}\star
P^{\dagger}\ .
\la{ai}
\ee

The algebra $hs_2(1)$ is a direct sum of spaces of monomials in $y$ and
$\yb$ of degree $2,6,10,...$. We use a notation such that if $P$
is an analytical function of $y$ and $\yb$ then

\be
P_{\a(k)\ad(l)}={1\over
k!l!}\del_{\a_1}\cdots\del_{\a_k}\bar{\del}_{\ad_1}\cdots
\bar{\del}_{\ad_l}P|_{Y=0}\ .
\la{not}
\ee

The space of bilinears of $hs_2(1)$ is isomorphic to $SO(3,2)$,
which is the maximal finite subalgebra of $hs_2(1)$.

From the above considerations it follows that $hs_2(1)$ can be
represented unitarily on $D(\ft12,0)$ and $D(1,\ft12)$. This
immediately yields a three-dimensional realization of $hs_2(1)$ as
a current algebra constructed from the singleton free field
theory. A four--dimensional field theory realization of $hs_2(1)$
must be based on a UIR of $hs_2(1)$ that decompose into $SO(3,2)$
UIR's with $E\geq s+1$. Such a UIR is given by the symmetric
tensor product

\be
\left(D(\ft12,0)\otimes  D(\ft12,0)\right)_S=
D(1,0)\oplus D(3,2) \oplus D(5,4) \cdots\ ,
\la{st}
\ee

which corresponds to a scalar, a graviton and a tower of massless
higher spin fields with spins $4,6,...$. Note that the spin
$s\leq 2$ sector of this spectrum contains a single real scalar
and as such it can not correspond to a bosonic subsector of a
(matter coupled) higher spin supergravity theory. Nonetheless the
scalar field is needed for unitary realization of the higher spin
$hs_2(1)$ symmetry and thus, even in the bosonic higher spin
theory, the field content is not arbitrary but rather is
restricted in an interesting way.

In fact, a spectrum of states with the same spin content can also
be obtained from the antisymmetric product of the fermionic
singletons $D(1,\ft12)$, the difference being that the scalar
field has lowest energy $E_0=2$ rather than $E_0=1$, which is the
case in \eq{st}.

It is worthwhile to note that the oscillator algebra has four
(linear) anti-involutions \cite{kv}: $\t$, $\t\pi$, $\t\bp$ and
$\t\pi\bp$, where $\pi$ and $\pb$ are defined below in \eq{p}.
Projecting by $\t$ and $\t\pi\pb$ leads to $hs_2(1)$, while
$\t\pi$ and $\t\bp$ leads to higher spin algebras that do not
contain the translations. The oscillator algebra also has an
involution $\rho\equiv\t^2=\pi\bp$. Projecting by imposing
$\r(P)=P$, gives rise to a reducible higher spin algebra $hs(1)$
\cite{kv} with spins $1,2,3,...$. The gauging of this algebra
gives rise to a spectrum given by $D(\ft12,0)\otimes D(\ft12,0)$.

To construct a four--dimensional field theory with symmetry
algebra $hs_2(1)$ and spectrum \eq{st} one needs to introduce an
$hs_2(1)$ valued gauge field $\cA_\m$ {\it and} a scalar master
field $\Phi$ in a representation of $hs_2(1)$ containing the
physical scalar $\f$, the spin $2$ Weyl tensor $C_{\a\b\c\d}$, its
higher spin generalizations $C_{\a(4n)}$ $n=2,...$
%%%%%%%%%%%%%%%%%%%%%%%%%%%%%%%%%%%%%%%%%%%%%%%%%%%%%%%%%%%%%%%%%%%%%%%%
\footnote {We use the notation $\a(n)\equiv (\a_1\cdots \a_n)$.
All spinor indices denoted by the same letter (with different
subscripts) are assumed to be symmetrized with unit strength.},
%%%%%%%%%%%%%%%%%%%%%%%%%%%%%%%%%%%%%%%%%%%%%%%%%%%%%%%%%%%%%%%%%%%%%%%%
and all the higher derivatives of these fields \cite{v}. Let us
first give an intuitive explanation of this, before we give the
formal construction of the theory. The dynamics for the gauge
fields follows from a curvature constraint of the form

\be
{\cal F}_{\m\n}=B_{\m\n}\ ,\qquad {\cal F}\equiv
d\cA+g\cA\star \cA
\la{c}
\ee

where two--form $B$ is a function of $\cA_\m$ and $\F$. It is
assumed that the structure of \eq{c} is analogous to \eq{weyl},
such that the Weyl tensors are given in terms of curvatures
\cite{vc}. In fact, it has been shown by Vasiliev (see \cite{vr}
for a review) that the spectrum \eq{st} requires that the
linearized expression for $B$ (in a $\F$ expansion) must obey

\be
B_{\a(m)\ad(n)}=\delta_{n0}e^a\wedge e^b
(\sigma_{ab})^{\b(2)}C_{\a(m-2)\b(2)}-h.c.\ ,\quad m+n=2,6,10,...\
\la{lc}
\ee

where we have expanded $B$, using the notation \eq{not}. From
\eq{c} it follows that $dB+g\cA\star B-gB\star \cA=0$, and there
are no further constraints. Thus, the master field $\F$ must
contain not only the scalar field and the Weyl tensors, but also
all their higher derivatives. The linearized scalar field equation
can also be written in first order form as follows:

\be
\del_{\a\ad}\f=ig\kappa^{-1}\F_{\a\ad}\ ,\quad
\na_{\a\ad}\F_{\b\bd}=ig\kappa^{-1}(\F_{\a\ad\b\bd}-\e_{\a\b}\e_{\ad\bd}\f)\ .
\la{se}
\ee

Thus we find that the scalar field $\f$, the generalized Weyl
tensors $C_{\a(4n)}$ and all their covariant derivatives fit into
$\F$ as follows:

\bea
\F|_{y=0} &=& \f\ ,
\nn\\
\F_{\a\ad}&=& -ig^{-1}\k\del_{\a\ad}\f+\cdots\ ,
\nn\\
\F_{\a(2)\ad(2)}&=&
i^2g^{-2}\k^2\del_{\a_1\ad_1}\del_{\a_2\ad_2}\f+\cdots\ ,
\nn\\
&\vdots& \nn\\[5pt]
\F_{\a(4n)}&=& g^{-2}\k^2C_{\a(4n)}\ ,
\nn\\
\F_{\a(4n)\b\bd}&=&  -ig^{-1}\k\del_{\b\bd} C_{\a(4n)}+\cdots\ ,
\nn\\
\F_{\a(4n)\b(2)\bd(2)} &=& i^2g^{-2}\k^2
\del_{\b_1\bd_1}\del_{\b_2\bd_2}C_{\a(4n)}+\cdots\ ,
\la{fc2}\\
&\vdots&
\nn
\eea

and the hermitian conjugates, where $n=1,2,\dots$, we have used
the notation in \eq{not} and the dots denote the
covariantizations. Thus, the nonzero components of $\F$ are
$\F_{\a(m)\ad(n)}$, $|m-n|=0$ mod $4$. This is equivalent to
imposing the conditions:

\be
\t(\F)=\bp(\F)\ ,\qquad \F^{\dagger}=\pi(\F)\ ,
\la{qa}
\ee

where

\be \pi(\F(y,\yb))=\F(-y,\yb)\ , \quad\quad
\pb(\F(y,\yb))=\F(y,-\yb)\ .\la{p} \ee

This defines a `quasi-adjoint' representation of $hs_2(1)$ with
covariant derivative

\be
{\cal D}\F=d\F+g\cA\star\F-g\F\star\pb(\cA)\ .
\la{cd}
\ee

Thus, the integrability condition on $B_2$ and the scalar field
equation \eq{se} must combine into a single constraint of the
form ${\cal D}_\m \F= B_\m$, where the one-form $B$ is a function
of $\cA_\m$ and $\Phi$. In summary, the higher spin field equations
are given by the constraints

\be\cF_{\m\n}=B_{\m\n}(\cA,\F)\ ,\qquad {\cal
D}_\m\F=B_\m(\cA,\F)\ .\la{b} \ee

%%%%%%%%%%%%%%%%%%%%%%%%%%%%%%%%%%%%%%%%%%%%%%%%%%%%%%%%%%%%%%%%%%%%

\section{Construction of the Constraints}

%%%%%%%%%%%%%%%%%%%%%%%%%%%%%%%%%%%%%%%%%%%%%%%%%%%%%%%%%%%%%%%%%%%%%

In order to construct the interactions in $B_{\m\n}$ and $B_\m$
one may employ a Noether procedure in which $d^2=0$ is satisfied
order by order in an expansion in $\F$ (counted by powers of
$g$). This can be facilitated by a geometrical construction based
on extending the ordinary four--dimensional spacetime by an
internal four--dimensional noncommutative space with spinorial
coordinates $z_\a$ and $\zb_{\ad}$ obeying the basic `contraction
rules'\cite{v,vr}

\bea
z_\a\star z_\b &=& z_\a z_\b - i \e_{\a\b}\ ,
\quad z_\a\star y_\b=z_\a y_\b+i\e_{\a\b}\ ,
\la{eosc1}\w2
y_\a\star z_\b &=& y_\a z_\b - i \e_{\a\b}\ ,\quad
y_\a\star y_\b=y_\a y_\b+i\e_{\a\b}\ ,
\la{eosc2}\w2
\zb_{\ad}\star \zb_{\bd} &=& \zb_{\ad}\zb_{\bd}-i\e_{\ad\bd}\ ,\quad
\zb_{\ad}\star\yb_{\bd}=\zb_{\ad}\yb_{\bd}-i\e_{\ad\bd}\ ,
\la{eosc3}\w2
\yb_{\ad}\star\zb_{\bd}&=&\yb_{\ad}\zb_{\bd}+i \e_{\ad\bd}\ ,\quad
\yb_{\ad}\star \yb_{\bd}=\yb_{\ad}\yb_{\bd}+i\e_{\ad\bd}\ ,
\la{eosc4}
\eea

together with $z\star \zb=z\zb$, $\zb\star z=\zb z$,
$z\star\yb=z\yb$ and $\zb\star y=\zb y$, with the following
generalization to arbitrary functions of $(y,\yb)$ and $(z,\zb)$:

\be
F\star G=\int d^4u~d^4v~
F(y+u,\yb+\uba;z+u,z-\uba)~G(y+v,\yb+\vb;z-v,\zb+\vb)~e^{i(u_\a
v^\a+\uba_{\ad}\vb^{\ad})}\ .
\ee

The above (associative) algebra is equivalent to the normal
ordered product of a pair of symplectic oscillators. In the
extended spacetime, one considers an integrable system consisting
of a one--form $\hcA=dx^\m\hcA_\m+dz^\a\hcA_\a+d\zb^{\ad}\hcA_{\ad}$
and scalar $\hF$ defined by \cite{vr}

\bea
\t(\hcA)&=&-\hcA\ ,\qquad\ \hcA^{\dagger}=\hcA\ , \la{ha}
\w2 \t(\hF)&=&\pb(\hF)\ ,\qquad \hF^{\dagger}=\pi(\hF)\ ,
\la{hf}
\eea

where the anti-involution $\t$ and the involutions $\pi$ and $\bp$
have been extended as follows:

\bea
\t(f(y,\yb;z,\zb))&=&f(iy,i\yb;-iz,-i\zb)\ ,\la{et}\w2
\pi(f(y,\yb;z,\zb))&=&f(-y,\yb;-z,\zb)\ ,\la{ep}\w2 \pb(f(y,\yb;z,\zb))&=&
f(y,-\yb;z,-\zb)\ .\la{epb}
\eea

By definition the exterior derivative $\hat{d}$ commutes with the
maps in \eqs{et}{epb}, such that $\t(dz^\a)=-i dz^\a$,
$\pi(dz^\a)=-dz^\a$ and $\pb(dz^\a)=dz^\a$.

The concise form of the full higher spin field equations
was first given in \cite{v}. As emphasized in \cite{ss1} these
equations are equivalent to the following curvature constraint:

\be
\hcF\equiv \hat{d}\hcA+g\hcA\star \hcA=\ft{i}4dz^\a\wedge
dz_\a \hF\star \kappa + \ft{i}4 d\zb^{\ad}\wedge d\zb_{\ad}
\hF\star \bar{\kappa}\ ,
\la{mc}
\ee

where the element $\kappa$ is defined by

\be
\kappa=\exp iz_\a y^\a\ ,\quad \bar{\kappa}\equiv \kappa^{\dagger}=\exp
-i\zb_{\ad} \yb^{\ad}\ .
\ee

Multiplication by $\kappa$ connects the two representations given
in \eqs{ha}{hf}, which follows from using \eq{ai} and the
following basic lemmas:

\be
\kappa\star f(y,\yb;z,\zb) =\kappa f(z,\yb;y,\zb)\ ,\qquad
f(y,\yb;z,\zb) \star \kappa = \kappa f(-z,\yb;-y,\zb)\ .\la{kappa}
\ee

This property is crucial for the whole construction, as we shall
see below. The Bianchi identity implies that $\F$ must satisfy

\be
{\hat{\cal D}} \hF \equiv \hat{d}\hF+g\hcA\star \hF-g\hF\star\pb(\hcA)=0\ .
\la{fc}
\ee

The rationale behind \eq{mc} is the following:

\begin{itemize}

\item[1)]

From the $\a\b$, $\a\bd$, $\a\m$ and $\ad\m$ components of
\eq{mc} and the $\a$ and $\ad$ components of \eq{fc}, we can
solve for the $z$ and $\zb$ dependence of all the fields in terms
of the `initial' conditions:

\be
\hcA_\m|_{Z=0}=\cA_\m\ ,\qquad
\hF|_{Z=0}=\F\ . \la{ic}
\ee

The integrability of \eq{mc} and \eq{fc} then implies that the
remaining $\m\n$ and $\m$  components of \eq{mc} and \eq{fc} are
satisfied for all $z_\a$ and $\zb_{\ad}$ provided that they are
satisfied at $z_\a=\zb_{\ad}=0$, that is

\be \hcF_{\mu\nu}|_{Z=0}=0\ ,\quad \qquad {\hat{\cal
D}}_\m\hF|_{Z=0}=0\ .\la{hse}\ee

These are equations of the form \eq{b}, which by construction
define an integrable set of constraints in ordinary spacetime.

\item[2)]

Upon linearizing the constraints \eq{mc} and \eq{fc}, it follows
that $\hF=\F(y,\yb)$. From \eq{kappa} it
then follows that

\be
\hF\star \kappa|_{Z=0}=\F(0,\yb)\ ,\ \qquad \hF\star {\bar
\kappa}|_{Z=0}=\F(y,0)\ .
\la{pure}
\ee

Comparing with \eq{fc2} one sees that the linearised two--form
$B_{\m\n}$ in \eq{b} depends on the the Weyl tensors $C_{\b(2s)}$,
$s=2,4,6,...$, but not on their derivatives, which is crucial in
order for the linearized field equations to be of the right form
\eq{lc}.

\item[3)]
Viewed as a Cartan integrable system, it is clear that Eqs.
\eq{mc} and \eq{fc} are gauge invariant (and that spacetime
diffeomorphisms are automatically incorporated into the gauge
group). The gauge transformations leaving \eq{mc} and \eq{fc}
invariant are

\be
\d \hcA=d\he +g\hcA\star \he - g\he \star \hcA\ ,\quad \d
\hF=g\he\star \hF-g\hF\star \pi(\hcA)\ .
\la{gt}
\ee

The $Z$--dependence in $\he$ can be used to impose the gauge
condition

\be
\hcA_\a|_{Z=0}=0\ .
\la{gc}
\ee

The gauge symmetries of the spacetime higher spin field equations
\eq{hse} then become

\be
\d \cA_\mu=\del_\m \e+g(\hcA_\m\star \he-\he\star
\hcA_\m)|_{Z=0}\ ,\quad \d \F= g(\he\star \hF-\hF\star
\pi(\he))|_{Z=0}\ ,
\label{rgt}
\ee

where the residual gauge transformations $\he$ are solved from

\be
\del_\a\he + g(\hcA_\a\star\he-\he\star \hcA_\a)|_{Z=0}=0\ ,
\quad \he|_{Z=0}=\e \ .
\ee

and its hermitian conjugate, where the initial condition $\e$
generates the original higher spin gauge algebra $hs_2(1)$.

\item[4)] Importantly, due to the mixing between $Y$ and $Z$ in
\eqs{eosc1}{eosc4}, both \eq{hse} and \eq{rgt} receive nontrivial
corrections which `deform' the spacetime curvatures such that
\eq{hse} describe an interacting system.

\end{itemize}

%%%%%%%%%%%%%%%%%%%%%%%%%%%%%%%%%%%%%%%%%%%%%%%%%%%%%%%%%%%%%%%%%%%%

\section{The Covariant Curvature Expansion}

%%%%%%%%%%%%%%%%%%%%%%%%%%%%%%%%%%%%%%%%%%%%%%%%%%%%%%%%%%%%%%%%%%%%

Provided that $g<<1$, it makes sense to solve the spinorial
components of the higher spin equations \eq{mc} and \eq{fc}
subject to the initial condition \eq{ic} by expanding in $\F$ as
follows:

\bea
\hF&=&\sum_{n=1}^{\infty} g^{n-1} \hF^{(n)}\ , \la{fe} \w2
\hcA_{\a}&=&\sum_{n=0}^{\infty} g^{n-1}\hcA_{\a}^{(n)} \ ,
\la{sce}\w2 \hcA_\m&=&\sum_{n=0}^{\infty} g^n\hcA^{(n)}_\m\ ,
\la{ce}
\eea

where the superscript $n$ refers to terms that are $n$'th order in
$\F$. We begin by expanding the purely spinorial components of
\eq{mc} as follows ($n\geq0$):

\bea
\del^\a\hcA^{(n)}_\a &=& \ft{i}2
\hF^{(n)}\star\kappa-\sum_{j=0}^{n} \hcA^{(j)\a}\star
\hcA_\a^{(n-j)}\ , \la{es2} \w2
\del_\a\hcA^{(n)}_{\ad}-\del_{\ad}\hcA^{(n)}_\a
&=&\sum_{j=1}^{n-1}\left(\hcA^{(j)}_{\ad}\star \hcA^{(n-j)}_\a -
\hcA^{(n-j)}_\a \star \hcA^{(j)}_{\ad}\right)\ ,
\la{es3}
\eea

and the spinorial components of \eq{fc} as follows ($n\geq 0$):

\be
\del_\a\hF^{(n)}=\sum_{j=1}^{n-1}\left(\hF^{(j)}\star
\pb(\hcA^{(n-j)}_\a) -\hcA^{(n-j)}_\a\star \hF^{(j)}\right)\ .
\la{es1}
\ee

These equations form an integrable equation system for
$\hcA^{(n)}_\a$ ($n\geq 0$) and $\hF^{(n)}$ ($n\geq 1$) subject
to the initial condition \eq{ic} that we solve by `zig-zagging'
back and forth between \eqs{es2}{es3} and \eq{es1}. For $n=0$, eqs.
\eqs{es2}{es3} show that
$\hcA^{(0)}_\a$ is a gauge artifact. In the gauge \eq{gc}, we can
therefore set

\be
\hcA^{(0)}_\a=0\ .
\ee

We continue the zig-zagging by taking $n=1$ in \eq{es1}, whose
right hand side also vanishes. The solution, satisfying the
initial condition \eq{ic} is therefore

\be
\hF^{(1)}=\F(y,\yb)\ .
\la{f1}
\ee

This result can then be used in \eqs{es2}{es3} for $n=1$ to solve
for $\hcA^{(1)}_\a$ as follows

\be
\hcA^{(1)}_\a=-\ft{i}2 z_\a \int_0^1 tdt~
\hF(-tz,\yb)\kappa(tz,y)\ .
\la{a1}
\ee

The results \eq{f1} and \eq{a1} can then be used in \eq{es1} for
$n=2$ to solve for $\hF^{(2)}$, and so on. This generates the
following series expansion $(n\geq 2)$:

\be
\hF^{(n)}= z^\a\sum_{j=1}^{n-1}\int_0^1 dt \left(
\hF^{(j)}\star \pb(\hcA^{(n-j)}_\a)-\hcA^{(n-j)}_\a\star
\hF^{(j)}\right) (tz,t\zb)\quad +\ {\rm h.c.}\ , \la{fn} \ee \bea
\hcA^{(n)}_\a&=&-z_\a \int_0^1 tdt\left(\ft{i}2 \hF^{(n)}\star\kappa
-\sum_{j=1}^{n-1}\hcA^{(j)\b}\star\hcA^{(n-j)}_\b\right)(tz,t\zb)
\nn\w2 &&+\zb^{\bd}\sum_{j=1}^{n-1}\int_0^1 tdt \left[
\hcA^{(j)}_{\bd},\hcA^{(n-j)}_\a\right](tz,t\zb)\ ,
\la{an}
\eea

where it is understood that the $\star$ products have to be
evaluated before replacing $(z,\zb)\ra(tz,t\zb)$ and that the
hermitian conjugate in \eq{fn} is in accordance with the reality
condition $\hF^\dagger=\pi(\hF)$.

Having analyzed the purely spinorial components of the constraint
\eq{mc}, we next analyze its $\a\m$ components, which give the
following equations for $\hcA^{(n)}_\m$ ($n\geq0$):

\be
\del_\a\hcA^{(n)}_\m=\del_\m\hcA^{(n)}_\a +\sum_{j=1}^n \left[
\hcA^{(n-j)}_\m, \hcA^{(j)}_\a\right]\ .
\ee

These equations, together with the initial condition \eq{ic}, can
be integrated to yield

\bea
\hcA^{(0)}_\m&=&\cA_\m\ ,\la{a0} \\
\hcA^{(n)}_\m&=&\int_0^1 dt\left[
z^\a\left(\del_\m\hcA^{(n)}_\a+\sum_{j=1}^n \left[
\hcA^{(n-j)}_\m, \hcA^{(j)}_\a\right]\right)(tz,t\zb)\right.
\nn\\
&&\qquad\,+\left.\zb^{\ad}\left(\del_\m\hcA^{(n)}_{\ad}+\sum_{j=1}^n
\left[ \hcA^{(n-j)}_\m,
\hcA^{(j)}_{\ad}\right]\right)(tz,t\zb)\right] \ .
\la{amn}
\eea

Note that in \eq{amn} the second line is minus the hermitian conjugate of the
first line. Substituting for $\hcA_\a^{(n)}$ and
$\hcA^{(n)}_{\ad}$ by the expression given in \eq{an}, we observe
that the first term in \eq{an} drops out, using $z^\a z_\a=0$.
The second term, which is real, cancels against the same term
coming from subtracting its hermitian conjugate, that is

\be
z^\a \hcA^{(n)}_\a + \zb^{\ad}\hcA^{(n)}_{\ad}=0\ .
\la{lemma}
\ee

This can be used to to simplify \eq{amn} further, with the result:

\bea
\hcA_\m^{(n)}&=&i\int_0^1{dt\over t}\left[\sum_{j=1}^{n}
\left(\hcA^{\a(j)}\star \del_\a^{(-)}\hcA^{(n-j)}_\m+
\del_\a^{(+)}\hcA^{(n-j)}_\m\star \hcA^{\a(j)}\right)\right.
\nn\\
&&\qquad \left.+\sum_{j=1}^{n}\left(\hcA^{\ad(j)}\star
\del_\a^{(+)}\hcA^{(n-j)}_\m+ \del_{\ad}^{(-)}\hcA^{(n-j)}_\m\star
\hcA^{\ad(j)}\right)\right](tz,t\zb)\ ,
\la{ann}
\eea

where $\del_\a^{(\pm)}=\del_\a^{(z)}\pm\del_\a^{(y)}$. Finally,
substituting \eq{f1}, \eq{fn}, \eq{a0} and \eq{ann} into the
spacetime components of \eq{mc} and \eq{fc} one obtains an
expansion of the constraints \eq{b} as follows:

\bea
{\cF}_{\m\n} &=&-\sum_{n=1}^\infty\sum_{j=0}^{n}
g^{n+1}\left( \hcA_{[\m}^{(j)}\star
\hcA_{\n]}^{(n-j)}\right)|_{_{Z=0}}\ ,
\la{cn}\w2
{\cal D}_\m \F &=&
\sum_{n=2}^\infty g^{n-1}\left( -\del_\m\hF^{(n)}|_{_{Z=0}} + g
\sum_{j=1}^{n} \left(\hF^{(j)}\star
\bp(\hcA^{(n-j)}_\m)-\hcA^{(n-j)}_\m\star \hF^{(j)}
\right)|_{Z=0}\right)\ ,
\la{bn}
\eea

where the unhatted curvature $\cF_{\m\n}$ and covariant
derivative ${\cal D}_\m\F$ are given in \eq{c} and \eq{cd}. The
following comments are in order:

\begin{itemize}

\item[1)]
By construction the constraints \eqs{cn}{bn} are integrable order by order
in $g$, with $hs_2(1)$ gauge symmetry given by \eq{rgt}.

\item[2)]
The curvatures $\cF_{\m\n,\a(s)\ad(s-2)}$ contain the dynamical
field equations for spin $s=2,4,6,...$. The remaining curvatures
are generalized torsion equations, except for the pure curvatures
$e_\m{}^a
e_\n{}^b(\s_{ab})_{(\a_1\a_2}\cF^{\m\n}{}_{,\a_3\dots\a_{2s-2})}$
which are set equal to the generalized Weyl tensors $\F_{\a(2s)}$.
The generalized torsion equations can be used to eliminate the
auxiliary gauge fields $\cA_{\m,\a(m)\ad(n)}$, $|m-n|\geq 2$ in
terms of the physical gauge fields $\cA_{\m,\a(s-1)\ad(s-1)}$.

\item[3)] The constraint \eq{bn} leads to the identifications
in \eq{fc2}, and thus in particular contains the full version of the scalar
field equation \eq{se}.

\item[4)] Setting $\cA_\m=\o_\m+W_\m$, where $\o_\m$
contain the $SO(3,2)$ gauge fields and $W_\m$ the higher spin fields,
the right hand side of \eq{cn} for $n=1$, that is $-2g^2
\cA^{(0)}_{[\m}\star \cA^{(1)}_{\n]}$, contains $\o^2\F$ terms of
the form \eq{lc}. Thus, at the linear in $\F$ level, we can consistently
set $W_\m=\f=0$, which then yields the full, covariant
Einstein equation \eq{ee} written in the form \eq{weyl}.

\end{itemize}

%%%%%%%%%%%%%%%%%%%%%%%%%%%%%%%%%%%%%%%%%%%%%%%%%%%%%%%%%%%%%%%%%%%%%%

\section{Comments}

%%%%%%%%%%%%%%%%%%%%%%%%%%%%%%%%%%%%%%%%%%%%%%%%%%%%%%%%%%%%%%%%%%%%%%

We have discussed how the constraint \eq{mc} in the extended
spacetime gives rise to a manifestly covariant expansion of the
higher spin field equations in terms of curvatures and a physical scalar
field. In particular we have found a significant simplification
in the interaction terms due to the identity \eq{lemma}. The
resulting higher spin equations contain the full Einstein equation
already at the leading order in the curvature expansion.

This raises the question of whether there exists any limit in
which the higher spin equations reduce to ordinary anti-de Sitter
gravity, possibly coupled to the scalar field $\f$.
There are two natural parameters to scale in the problem: the
gauge coupling $g$ and a 'noncommutativity' parameter $\l$ that
can be introduced by taking $y\ra \l y$ and $z\ra \l z$.
Scaling $g$ is equivalent to scaling the derivatives $\del_\m$,
that is to taking a low energy, or weak curvature, limit,
while scaling $\l$ is equivalent to taking a Poisson limit of the
higher spin algebra. In the latter limit the multicontractions are
suppressed such that the commutator of a spin $s$ and a spin $s'$
generator closes on a spin $s+s'-2$ generator.

One important issue to settle is whether the scalar $\f$
disappears in this limit or if it stays in the spectrum, in which case
the scalar potential should be computed from the expansion given above
(in which case the term curvature expansion would of course strictly speaking
be inappropriate). This issue is of extra interest in cases with extended
supersymmetry since gauged supergravity requires scalars in particular cosets
that determines the potential. In the nonsupersymmetric case a nontrivial
potential could also be of interest, since it may cause the scalar to
flow to other vacua than the anti-de Sitter vacuum at $\f=0$.

Other points of interest are, that in analogy with results for
$W$-gravity in two dimensions \cite{wg}, the Poisson limit could
be considered as a mean for extracting a 'classical' higher spin
theory whose quantization would then yield the full higher spin
theory discussed here. There is also the issue of whether one can
organize the expansion in curvatures such that it can be compared
with the $\alpha'$ corrections coming from string theory, or
M-theory. We will report elsewhere on some of the topics discussed
above \cite{ss2}.

\medskip

%%%%%%%%%%%%%%%%%%%%%%%%%%%%%%%%%%%%%%%%%%%%%%%%%%%%%%%%%%%%%%%%%%%%%%%

\noindent{\large \bf Acknowledments}

%%%%%%%%%%%%%%%%%%%%%%%%%%%%%%%%%%%%%%%%%%%%%%%%%%%%%%%%%%%%%%%%%%%%%%%

One of the authors (E.S.) wishes to thank I. Bars and E. Witten
for stimulating discussions. This research has been supported in
part by NSF Grant PHY-0070964. P.S. is supported by Stichting FOM,
Netherlands.

%%%%%%%%%%%%%%%%%%%%%%%%%%%%%%%%%%%%%%%%%%%%%%%%%%%%%%%%%%%%%%%%%%%%%%

\ed